\documentclass[aps,pra,twocolumn,10pt,showpacs]{revtex4-1}

\usepackage{amsmath}
\usepackage{graphicx}
\usepackage{ulem}
\usepackage{times}
\usepackage{amssymb}
\usepackage{hyperref}
\usepackage{textcomp}
\usepackage{xcolor}
\usepackage[T1]{fontenc}
\usepackage[margin=2cm]{geometry}

\newcommand{\tno} {\,^\circ_\circ\,}

\begin{document}

\title{Quantum-entangled light from localized emitters}
\author{P. Gr\"unwald}
\email{Electronic address: peter.gruenwald@uni-rostock.de}
\author{W. Vogel}
\affiliation{Arbeitsgruppe Theoretische Quantenoptik, Institut f\"ur Physik, Universit\"at Rostock, D-18055 Rostock, Germany}
\date{\today} 

\begin{abstract}
Localized radiation sources are analyzed with respect to the relation of nonclassicality and quantum entanglement of the emitted light. The source field parts of the radiation emitted in different directions are closely related to each other. As a consequence, nonclassicality of the light fields in one direction directly implies entanglement of the field modes in different directions. This implication can be extended to multipartite-entanglement and multi-time quantum correlations. Given that a nonclassical effect is observed, our approach explicitly yields the multipartite entanglement witnesses.
Two examples are considered, the fluorescence radiation of a system of two-level atoms and of excitons in a semiconductor quantum well.
\end{abstract}


\maketitle

\section{Introduction}
Nonclassical light fields possess features which cannot be explained by Maxwell's field theory. 
More precisely, a light field is nonclassical if the $P$~function of Glauber and Sudarshan~\cite{Gl,Su} cannot be interpreted as a classical probability density~\cite{Ti,Ma}. A special nonclassical phenomenon is quantum entanglement~\cite{We}, with quantum correlations between separated subsystems. Entanglement is considered as one of the major resources for the development of future quantum technologies~\cite{QuEnt}. However, the relation of entanglement to certain properties of the $P$~function is yet unknown. Consequently, the general relation between nonclassicality and quantum entanglement is still an open problem of great interest. 

Two entangled radiation modes propagating in different directions can be obtained by splitting a nonclassical input field by a beam splitter~\cite{Aharonov,OHM,Kim02,Wang02}. Based on a unified quantification of nonclassicality and entanglement, it has been shown that a beam splitter even transforms a single-mode nonclassical state of a given amount of nonclassicality into the same amount of bipartite entanglement~\cite{Vo-Sp14}. More generally, an $N$ splitter creates genuine multipartite entanglement of the same amount as the nonclassicality of the single-mode input state. Even without using a beam splitter, a sample of noninteracting atoms was shown to emit entangled light in different directions~\cite{PRL2010} if the light field in a single direction has a sub-Poisson photon statistics.

Based on observable moments of photon annihilation and creation operators of arbitrarily high orders~\cite{Ag-Ta,Ag}, the nonclassical properties of single-mode radiation fields can be fully characterized~\cite{EvgenyNC05}. Quantum entanglement can also be characterized in terms of moments~\cite{EvgenyEnt05}, for states with a negative partial transposition (NPT). This approach unifies a variety of second-order~\cite{Simon,Duan} and higher-order entanglement criteria~\cite{ArgaBis,HillZub}. Note that NPT entanglement is an important class of quantum entanglement, which is useful for many applications~\cite{PPT-Theo2}. Based on the methods in~\cite{EvgenyNC05,EvgenyEnt05}, the relation between nonclassicality and entanglement conditions has been discussed in some detail~\cite{Mira2010}. 

The characterization by moments has also been generalized to genuine multipartite NPT entanglement~\cite{EvgenyEnt06}. On the other hand, nonclassicality conditions have been introduced which characterize general space-time-dependent nonclassical correlation properties in terms of normal- and time-ordered field correlation functions~\cite{Vogel08}. These conditions are fundamental for the proper characterization of the nonclassicality of radiation fields emitted by atomic sources. All the moments and correlation functions needed in this context can be observed by homodyne correlation measurements~\cite{EvgenyMeas06}.

In this contribution we study the relation between the nonclassicality of light emitted by a localized (small compared with the emission wavelength) source and the quantum entanglement of the fields observed in different directions. Based on the nonclassicality and entanglement criteria in terms of moments, we prove that nonclassical emission in a single direction implies NPT entanglement of the fields in different directions. The extension to multipartite NPT entanglement and even multitime quantum correlations is also given. For any observed nonclassical effect, the corresponding multipartite entanglement witnesses follow directly. We will briefly consider the examples of the fluorescence of a sample of $N$ atoms and of excitons in a quantum well.

The paper is organized as follows. In Sec.~II we reconsider the known criteria for nonclassicality and NPT entanglement in terms of field moments. The relation between nonclassicality and entanglement conditions is analyzed in Sec.~III for a localized radiation source. In this context, both bipartite and multipartite entanglement are studied and examples of typical light sources are discussed. In Sec.~IV we extend the treatment to general space-time-dependent quantum correlations of the light emitted by the sources under study. A summary and some conclusions are given in Sec.~V.

\section{Nonclassicality and entanglement conditions}
Let us consider a light field $\hat{\vec E}(\vec r,t)$ emitted from an arbitrary light source and detected at space-time point $\vec r,t$. The extension of the source shall be small compared to the central wavelength, so that retardation effects are included in $t$ and we omit the time arguments for single-time fields and expectation values in the following. The field consists of a source field part $\hat{\vec E}_\text s$, the actual emission from the source, and a free field part $\hat{\vec E}_\text f$. If a given function of field operators is normally and time ordered~\cite{WelVo}, and the free fields are in the vacuum state at the detector, the latter do not contribute to the expectation value of the function of field operators.

The vector character of the fields is encoded within the spatial modes. The detector measures the projection onto the direction from the source to the detector, so we consider only the modulus. Both free and source field operators split into positive- and negative-frequency parts
\begin{equation}
   \hat E_\text{f,s}=\hat E_\text{f,s}^{(+)}+\hat E_\text{f,s}^{(-)}.
\end{equation}

With the above description, we may now formulate nonclassicality conditions for the fields according to~\cite{EvgenyNC05,Vogel08}. Consider an operator function $\hat h$ expanded in terms of the fields as
\begin{equation}
   \hat h=\sum_{n,\ell=0}^\infty c_{n,\ell}\hat E^{(-)n}\hat E^{(+)\ell},\label{eq.hNC}
\end{equation}
with $\hat E^{(\pm)}=\hat E_\text{f}^{(\pm)} + \hat E_\text{s}^{(\pm)}$. Assuming that the expectation values of the moments of the fields exist, the expectation value
\begin{equation}
   \langle\hat h^\dagger\hat h\rangle=\text{Tr}\{\hat h^\dagger\hat h\hat\varrho\}
\end{equation}
is non-negative for all quantum states $\hat\varrho$. Nonclassicality is verified if a normally ordered form fulfills the condition
\begin{equation}
   \langle:\hat h^\dagger\hat h:\rangle=\text{Tr}\{:\hat h^\dagger\hat h:\hat\varrho\}<0.\label{eq:ncl-cond}
\end{equation}
Herein $:\cdots:$ denotes normal ordering, i.e., the negative-frequency operators are ordered to the left of the positive-frequency operators~\cite{WelVo}. This general condition yields an infinite hierarchy of conditions for minors of the matrix of normal-ordered field moments; compare~\cite{EvgenyNC05}. If one of these minors is negative, nonclassicality occurs. The other way around, if all minors are positive, the state is classical. Note, that the normal ordering of all correlations allows us to omit the free fields in the calculations. 

Now consider two emitted fields $\hat E_1$ and $\hat E_2$. Quantum entanglement of the two fields occurs if the combined state for both fields fails to be in a separable form~\cite{We}. A specific entanglement condition is the Peres-Horodecki criterion~\cite{Peres,Horodecki96}, which states that a separable quantum state must remain a valid quantum state under partial transposition. 
On this basis, entanglement conditions have been formulated in terms of moments~\cite{EvgenyEnt05}. 
Consider the operator function $\hat f$ of the fields,
\begin{equation}
   \hat f=\sum_{m,n,k,\ell=0}^\infty c_{m,n,k,\ell}\hat E^{(-)m}_2\hat E^{(+)n}_2\hat E^{(-)k}_1\hat E^{(+)\ell}_1\label{eq.fEnt}
\end{equation}
with complex coefficients $c_{m,n,k,\ell}$. According to the Peres-Horodecki criterion, a state $\hat\varrho$ is entangled, if a partially transposed form fulfills the condition
\begin{equation}
   \langle(\hat f^\dagger\hat f)^\text{PT}\rangle=\text{Tr}\{(\hat f^\dagger\hat f)^\text{PT}\hat\varrho\}<0, \label{eq:NPT-ent}
\end{equation}
where ``PT'' stands for partial transposition. The transposition acts solely on the field components with index ``2'' with $(\hat E^{(\pm)}_2)^\text T=\hat E^{(\mp)*}_2$. Besides the phase factor, which is irrelevant for our discussion, transposition replaces the fields by the adjoint ones. This approach resembles that for nonclassicality and it yields a corresponding hierarchy of conditions for the minors of matrices of moments. If one of those minors is negative, the state is called NPT entangled. Note that also entanglement with positive partial transposition exists~\cite{PPT-Theo1,PPT-Theo2,PPT-Exp1,PPT-Exp2}. Transposition does not automatically yield a normally ordered structure, so free fields may be relevant. However, if a combination of field operators is already normally ordered, its transposed form is also normally ordered, as
\begin{equation}
   \left[\hat E^{(-)n}\hat E^{(+)k}\right]^\text T=\left(\hat E^{(-)k}\hat E^{(+)n}\right)^*
\end{equation}
holds for all $n,k\in\mathbb N$.

\section{Nonclassicality implies entanglement}
We are interested in the quantum correlations of the fields propagating from the same source in two different directions. For this purpose we have to analyze the source field $\hat E_\text s$.
The source field of a localized emitter can be written as
\begin{equation}
   \hat E_\text s^{(+)}=\chi(\vec r)\hat S,\quad\hat E_\text s^{(-)}=\chi^*(\vec r)\hat S^\dagger\label{eq.sourcefields}
\end{equation}
where $\chi(\vec r)$ is the spatial mode function, while $\hat S$ and $\hat S^\dagger$ represent the source operators. For the derivation of Eq.~(\ref{eq.sourcefields}) we refer to the Appendix. We made use of the localization of the source together with a small bandwidth of the emission spectrum compared with its central emission frequency~\cite{WelVo}. For two propagation directions ($j=1,2$), the difference of the two spatial field modes is encoded solely in the mode functions $\chi_j(\vec r)$. The operator structure is the same, as pointed out in~\cite{PRL2010}. At this point the hierarchy structure of equations for nonclassicality and entanglement becomes advantageous. In every minor, which can be calculated, every summand has the same number of field operators $\hat E_j^{(\pm)}$ and therefore the same number of prefactors $\chi_j(\vec r)$. Therefore, if the moments in $(\hat f^\dagger\hat f)^\text{PT}$ are normally ordered, we can extract the mode functions as prefactors, and the entanglement hierarchy reduces to a hierarchy of normally ordered minors of the source field operators $\hat S$. Such NPT-entanglement conditions for the light propagating in different directions from a single source exactly correspond to the nonclassicality conditions in a single propagation direction.

\subsection{Bipartite entanglement}
We need functions $\hat f$ for two-mode fields, for which $(\hat f^\dagger\hat f)^\text{PT}$  is normally ordered. For a general function $\hat f$ as defined in Eq.~(\ref{eq.fEnt}), $\hat f^\dagger\hat f$ has eight different field operators with different powers of positive- and negative-frequency parts. However, normal ordering requires all negative-frequency operators to be on the left of all positive-frequency operators. Hence, no more than two different field operators can be present in $\hat f$. Choosing $m=k=0$ in Eq.~(\ref{eq.fEnt}), we obtain
\begin{align}
   \hat f=&\sum_{n,\ell=0}^\infty c_{n,\ell}\hat E^{(+)n}_2\hat E^{(+)\ell}_1\label{eq.fentsol}\\
   \hat f^\dagger\hat f=&\hspace{-0.3cm}\sum_{n,n',\ell,\ell'=0}^\infty\hspace{-0.3cm} c^*_{n',\ell'}c_{n,\ell}\hat E^{(-)\ell'}_1\hat E^{(-)n'}_2\hat E^{(+)n}_2\hat E^{(+)\ell}_1\\
   (\hat f^\dagger\hat f)^\text{PT}=&\hspace{-0.3cm}\sum_{n,n',\ell,\ell'=0}^\infty\hspace{-0.3cm} c^*_{n,\ell'}c_{n',\ell}\hat E^{(-)\ell'}_1\left(\hat E^{(-)n'}_2\hat E^{(+)n}_2\right)^\text T\hat E^{(+)\ell}_1\nonumber\\
   =&\hspace{-0.3cm}\sum_{n,n',\ell,\ell'=0}^\infty\hspace{-0.3cm} c^*_{n,\ell'}c_{n',\ell}\hat E^{(-)\ell'}_1\left(\hat E^{(-)n}_2\hat E^{(+)n'}_2\right)^*\hat E^{(+)\ell}_1. \label{eq:f^+f-PT}
\end{align}
The transposition leaves the operator structure normally ordered, as $\hat E^{(-)n'}_2\hat E^{(+)n}_2$ is normally ordered. Hence, any NPT-entanglement condition~(\ref{eq:NPT-ent}) based on $\hat f$ in Eq.~(\ref{eq.fentsol}) corresponds to a nonclassicality condition of the type in Eq.~(\ref{eq:ncl-cond}).

The function $\hat f^\dagger\hat f$ includes positive- and negative-fre\-quency operators of the fields in two propagation directions, $\hat E_{1/2}^{(\pm)}$, occurring in different powers and in normal ordering. The corresponding nonclassicality conditions must also contain negative and positive-frequency field operators. Therefore, these conditions are given for the function
\begin{align}
    \hat h=&\sum\limits_{n,\ell=0}^\infty c_{n,\ell}\hat E^{(-)n}\hat E^{(+)\ell}.\label{eq.hentsol}
\end{align}
This is identical to Eq.~(\ref{eq.hNC}), which yields a complete characterization of nonclassicality; cf.~Eq.~(\ref{eq:ncl-cond}). Hence, we conclude that, for any localized source of nonclassical light, the light fields in different directions are NPT entangled.

Before we generalize this result to the multipartite case, we want to stress the necessity of the locality condition. For extended light sources, the quantum characteristics of the fields propagating in different directions are different in general. Therefore, nonclassicality observed in one direction is irrelevant for the entanglement of fields observed in different directions. Only for a localized source can the field operators in each propagation direction be related to the source operators in a unique way; for details see the Appendix.

\subsection{Multipartite entanglement}
The extension of the bipartite NPT-entanglement criteria to multimode fields was developed in~\cite{EvgenyEnt06}. The case of multipartite criteria based on second-order moments was studied earlier~\cite{vanLoock}; for its experimental application, see~\cite{Aoki}. Consider a system of  $M$ spatial field modes in different propagation directions from the source, with positive frequency field operators $\hat E_j^{(+)}$, $j=1,\ldots, M$. In this case the general function $\hat f$ of the fields can be written as a product combination of positive- and negative-frequency parts of all field modes,
\begin{align}
   \hat f=&\sum_{\mathbf{k,l}=0}^\infty c_\mathbf{k l}\mathbf{\left(\hat E^{(-)}\right)^k}\mathbf{\left(\hat E^{(+)}\right)^l}\label{eq:multipartite},\\
   \intertext{with}
   \mathbf k=&(k_1,\ldots,k_M),\ \mathbf l=(l_1,\ldots,l_M),\\
   \mathbf{\left(\hat E^{(-)}\right)^k}=&\prod_{j=1}^M\left(\hat E_j^{(-)}\right)^{k_j}.
\end{align}
Now, the partial transposition of $m$ field modes ($0<m<M$) in $\hat f^\dagger\hat f$ is performed. It reveals partial entanglement of the subsystems of the $m$ transposed and the $(M-m)$ untransposed modes via a similar hierarchy of minors of moments as in the bipartite case. We choose $\mathbf k=0$ in Eq.~(\ref{eq:multipartite}). Without loss of generality, we assume that the $m$ field modes to be transposed are ordered to the left of the $M-m$ untransposed field modes in operator $\hat f$. In this way, the action of partial transposition for any bipartite splitting yields a direct generalization of the case of bipartite entanglement,
\begin{align}
   \hat f^\dagger\hat f=&\sum_{\mathbf {l',l}=0}^\infty c_\mathbf{l'}^*c_\mathbf{l}\mathbf{\left(\hat E^{(-)}\right)^{l'}}\mathbf{\left(\hat E^{(+)}\right)^l},\\
   (\hat f^\dagger\hat f)^\text{PT} =&\sum_{\mathbf {l',l}=0}^\infty c_\mathbf{\tilde l'}^*c_\mathbf{\tilde l}\left(\hat E^{(-)}_M\right)^{l'_M}\cdots\left(\hat E^{(-)}_{m+1}\right)^{l'_{m+1}}\nonumber\\
   &\hspace*{1cm}\times\left[\left(\hat E^{(-)}_m\right)^{l'_m}\cdots\left(\hat E^{(+)}_{m}\right)^{l_{m}} \right]^\text{T}\nonumber\\
   &\hspace*{1cm}\times\left(\hat E^{(+)}_{m+1}\right)^{l_{m+1}}\cdots\left(\hat E^{(+)}_{M}\right)^{l_{M}}\nonumber\\
   =&\sum_{\mathbf {l',l}=0}^\infty c_\mathbf{\tilde l'}^*c_\mathbf{\tilde l}\left(\hat E^{(-)}_M\right)^{l'_M}\cdots\left(\hat E^{(-)}_{m+1}\right)^{l'_{m+1}}\nonumber\\
   &\hspace*{1cm}\times\left(\hat E^{(-)*}_m\right)^{l_m}\cdots\left(\hat E^{(+)*}_{m}\right)^{l'_{m}} \nonumber\\
   &\hspace*{1cm}\times\left(\hat E^{(+)}_{m+1}\right)^{l_{m+1}}\cdots\left(\hat E^{(+)}_{M}\right)^{l_{M}}. \label{eq:f^+f-mult}
\end{align}
The tilde on the multi-indices  indicates the transposition of the corresponding operators. The partial transposition is now normally ordered, and for a single source field, the entanglement conditions lead to the corresponding nonclassicality conditions, as in the bipartite scenario. A few things should be noted. The partial entanglement in an $M$-mode system requires at least an $M$th-order nonclassicality condition to be fulfilled. In turn, the proof of higher-than-bipartite cases in experiments also requires measuring these higher order correlations~\cite{EvgenyMeas06,Blatt}. On the other hand, the choice of the partition of the two subsystems is irrelevant for the argumentation. Thus, our findings are general enough for us to conclude, that one fulfilled $M$th-order nonclassicality condition verifies NPT-entanglement for any bipartite splitting of the $M$ fields propagating in $M$ directions, and hence genuine multipartite entanglement.

At this point it is worth noting that our theory not only yields the strict relation between single-mode nonclassicality and quantum entanglement for the light sources under study. In addition, we also provide the explicit form of the nonclassicality and entanglement witnesses~\cite{QuEnt}. Let us assume one has detected a particular nonclassical effect in a single propagation direction. This effect is related to a nonclassicality witness, by combining Eq.~(\ref{eq.hNC}) for properly chosen coefficients with (\ref{eq:ncl-cond}). The bipartite and multipartite entanglement witnesses directly follow by combining Eq.~(\ref{eq:NPT-ent}) with (\ref{eq:f^+f-PT}) and (\ref{eq:f^+f-mult}), respectively. Applying balanced homodyne correlation measurements~\cite{EvgenyMeas06}, this yields the full method for the verification of the kinds of entanglement we are interested in.

\subsection{Typical examples of light sources}
Let us consider two examples to illustrate our theory. In the first case, we analyze the resonance fluorescence of a sample of $N$ identical two-level atoms, each described by ground and excited states $|1 \rangle^{(k)}$ and $|2\rangle^{(k)}$, respectively, with $k=1,\ldots,N$. The source field operator $\hat E^{(+)}$ is given by the sum of the atomic lowering operators $\hat A_{12}^{(k)}=|1\rangle^{(k)}\,^{(k)}\langle 2|$:
\begin{equation}
   \hat E^{(+)}=\chi(\vec r)\sum_{k=1}^N\hat A_{12}^{(k)}e^{i\phi_k},
\end{equation}
where $\phi_k$ is the individual phase of the $k$th atom, including the laser phase, as well as the different positions of the atoms with respect to each other. Keep in mind that we still require a localized structure. Hence, we can assume that the atoms emit in cooperative manner~\cite{Dicke, Superfluor}. For $N<\infty$, the light source is always nonclassical, as each atom is a single-photon emitter, and hence only $N$ photons can be emitted at the same time. This is equivalent to the argument of photon antibunching in single-atom fluorescence, and this nonclassicality follows from the condition
\begin{equation}
   \hat h=c_{0,0}+c_{k,N+1-k}\hat E^{(-)k}\hat E^{(+)N+1-k}, \label{eq.atomNC}
\end{equation}
with $k\in\{1,\ldots, N\}$. In $:\hat h^\dagger\hat h:$, the term $\hat E^{(-)N+1}\hat E^{(+)N+1}$ is zero, resulting in a negative value for the corresponding minor if $\langle\hat E^{(-)k}\hat E^{(+)N+1-k}\rangle\neq0$. The corresponding entanglement function reads as
\begin{align}
    \hat f=c_{0,0}+c_{k,N+1-k}\hat E_2^{(+)k}\hat E_1^{(+)N+1-k}.\label{eq.atomEnt}
\end{align}
Therefore the different field modes of a sample of atoms always show entanglement. In~\cite{PRL2010}, it was already argued that sub-Poisson light from such a source yields entanglement in different directions. Now we can generalize this statement. Independent of the number and configuration of the atoms, there will always be a nonclassicality condition fulfilled and thus quantum entanglement of the fields in different directions can be concluded to occur. For a moderate number $N$ of atoms, Eq.~(\ref{eq.atomEnt}) always gives a reasonable criterion to detect entanglement. For example, in~\cite{PRL2010}, we could even consider the case of $N=1$. However, for $N\gg1$, both nonclassicality and entanglement may be difficult to detect. It should therefore be noted that other, lower-order correlation functions may also yield entanglement for certain regimes (e.g. squeezing for lower intensities and bistability, see~\cite{Raizen}).

The second system is the steady-state cooperative fluorescence of a sample of excitons in a GaAs quantum well~\cite{PRB2013,QW-Corr}. The excitons can be described as one collective bosonic excitation $\hat A$, with a Kerr-nonlinearity. However, as the material must first absorb photons to excite the excitons, the source field is also scaled with the absorption spectrum of the medium. The time-dependent fields become a convolution between the exciton operators and the absorption spectrum. In~\cite{QW-Corr}, we developed an algorithm to determine correlation functions of arbitrary  field operators to study the nonclassical properties of the quantum-well emission fields. For sufficiently low pump intensities the emission is squeezed, while for higher intensities we find sub-Poisson photon statistics. Both nonclassical effects, squeezing and sub-Poisson statistics, follow for the choices
\begin{align}
   \hat h_\text{Sq}=&c_{0,0}+c_{1,0}\hat E^{(-)}+c_{0,1}\hat E^{(+)}
\intertext{and}
   \hat h_\text{sP}=&c_{0,0}+c_{1,1}\hat E^{(-)}\hat E^{(+)},
\end{align}
respectively. Hence, the excitons emit quantum entangled light fields in different directions. 

\section{Multi-time quantum correlations}
Nonclassicality is not only described in terms of single-time correlation functions. Photon antibunching, first demonstrated in the resonance fluorescence of an atomic beam~\cite{ki-dag-ma}, was based on the correlation between the field intensities at different time points. For the study of multitime nonclassicality of radiation fields, the theory of nonclassicality was generalized in terms of normal- and time-ordered correlation functions~\cite{Vogel08}. The main essence of this extension is the inclusion of time ordering in the structure, that is, the negative-frequency operators are ordered with increasing time arguments from left to right, and the positive-frequency operators the other way around. As an example, consider the intensity-intensity correlation for a stationary field
\begin{equation}
\begin{split}
   G^{(2)}(\tau)=&\langle\tno\hat I(0)\hat I(\tau)\tno\rangle\\
   =&\langle\hat E^{(-)}(0)\hat E^{(-)}(\tau)\hat E^{(+)}(\tau)\hat E^{(+)}(0)\rangle.\label{eq.G2def}
\end{split}
\end{equation}
Herein $\hat I$ is the intensity, we assumed $\tau\geq0$, and $\tno\cdots\tno$ denotes time and normal ordering. If $G^{(2)}(\tau)>G^{(2)}(0)$, the light is antibunched. The formulation of general space-time-dependent nonclassicality criteria was based  on the expansion of operator functions $\hat h$ into a power series of field operators at different space-time points. As was the case in the previous discussions, for any classical light field, $\langle\tno\hat h^\dagger\hat h\tno\rangle$ is positive semidefinite, and any negativity is a signature of quantum correlation effects. The general structure of $\langle\tno\hat h^\dagger\hat h\tno\rangle$ again leads to a hierarchy of inequalities for various types of correlation functions.

At this point we can again use the fact that the transposition preserves the ordering of a time- and normal-ordered operator structure. Consider two fields propagating in different directions and at different times.  With the choice
\begin{equation}
   \hat f=\sum_{n,\ell=0}^\infty c_{n,\ell}\hat E^{(+)n}_2(t+\tau)\hat E^{(+)\ell}_1(t),
\end{equation}
with $\tau\geq0$, the partial transposition is again well ordered. Combining both multipartite and multi-time description is also straightforward. For the function
\begin{equation}
   \hat f=\mathbf{\left(\hat E^{(+)}(t)\right)^k}=\prod_{j=1}^M\left(\hat E_j^{(+)}(t_j)\right)^{k_j}\label{eq.multi-time-mode}
\end{equation}
with $t_j\leq t_\ell$ for $j>\ell$, the partial transposed form of the overall operator structure is well ordered, and the argument for omitting free fields persists. In this way, we obtain that all nonclassical correlation conditions discussed in~\cite{Vogel08} correspond to multipartite multitime NPT-entanglement conditions for a pointlike light source. It has to be stressed that timelike entanglement is a strongly debated topic and many questions concerning detection and interpretation of this phenomenon are still open~\cite{TimeEnt1,TimeEnt2,TimeEnt3}. Therefore, we avoid further interpretation of this result; it just yields the extension of multipartite entanglement to different times.

\section{Summary and Conclusions}

We have studied the relation between the nonclassicality of light from a localized radiation source and quantum entanglement of the fields emitted in different directions from the source. For localized sources, the operator structure of the fields propagating in different directions is uniquely related to the source operators. The only difference between different directions is in the prefactors representing the spatial modes. Therefore, an equivalence exists between the nonclassicality and NPT-entanglement criteria formulated in terms of radiation-field moments. From this equivalence it follows that all sources of nonclassical light, which are localized in a region smaller than the wavelength, emit light that is entangled in different directions. This conclusion is not valid for extended sources, where the nonclassicality of the light emitted in one direction does not imply quantum entanglement between the light propagating in different directions.

The equivalence between nonclassicality and entanglement can be extended to multipartite entanglement. We showed that a properly chosen nonclassicality condition of the field in one direction can even identify genuine multipartite entanglement between the fields in various directions. Furthermore, our method also provides the desired entanglement witnesses, which are based on established measurement techniques.
Two examples have been given to illustrate our theory, the fluorescent emission of both an ensemble of two-level atoms and of an exciton spot in a semiconductor quantum well. Finally, our results can also be extended to general multitime quantum correlations of light. 

\acknowledgments
The authors gratefully acknowledge support by the Deutsche Forschungsgemeinschaft through SFB 652 and valuable comments by Jan Sperling.

\section*{Appendix: Derivation of Eq.~(\ref{eq.sourcefields})}
We follow the definitions of~\cite{WelVo}. For a system of atomic emitters at the different positions $\vec r_n$, the polarization is given by
\begin{equation}
   \hat{\vec P}(\vec r)=\sum\limits_n\hat{\vec d}_n\delta(\vec r-\vec r_n),
\end{equation}
with $\hat{\vec d}_n$ being the dipole operator of emitter $n$. For a localized structure, the variation of the positions $\vec r_n$ is very small, so that they can be identified with the source position, $\vec r_n\approx\vec r_\text s$. Thus, we can take the $\delta$ function out of the sum and obtain
\begin{equation}
   \hat{\vec P}(\vec r)\approx\delta(\vec r-\vec r_\text s)\sum\limits_n\hat{\vec d}_n=\hat{\vec C}\delta(\vec r-\vec r_\text s). \label{eq.pol}
\end{equation}
In the Heisenberg-picture, the dipole operators become time dependent, $\hat{\vec C}\rightarrow\hat{\vec C}(t)$.

The source-field annihilation operator $\hat a_{\lambda,\text s}(t)$ for a single mode $\lambda$ is given by
\begin{equation}
\begin{split}
	\hat a_{\lambda,\text s}(t)=\frac{\omega_\lambda}{\hbar}&\int dt'\,\theta(t-t')\\
	&\times\int d^3\vec r'\,\vec A_\lambda(\vec r')\cdot\hat{\vec P}(\vec r',t')\,e^{-i\omega_\lambda(t-t')}
	\label{eq.sourceexpl}
\end{split}
\end{equation}
and the full positive-frequency source field follows as
\begin{align}
   \hat{\vec E}_\text s^{(+)}(\vec r,t)=&\sum_\lambda\vec F_\lambda(\vec r)\hat a_{\lambda,\text s}(t).
\end{align}
Applying Eq.~(\ref{eq.pol}) in~(\ref{eq.sourceexpl}), the $\vec r'$ integration can be solved. In the small-bandwidth limit, where for different $\lambda\neq\lambda'$ we have
\begin{equation}
   \left|\frac{\omega_\lambda-\omega_{\lambda'}}{\omega_\lambda}\right|\ll1,
\end{equation}
we can equate the frequencies with a characteristic source frequency, $\omega_\lambda=\omega_\text s$. Furthermore, $\vec A_\lambda(\vec r_\text s)\approx\vec A_{\rm s} (\vec r_\text s)$ is almost constant for all relevant modes $\lambda$. The overall source field  may then be split into parts as
\begin{align}
   \hat{\vec E}_\text s^{(+)}(\vec r,t)&=\vec\chi(\vec r)\hat{S}(t),\\
   \vec\chi(\vec r)&=\sum\limits_\lambda\vec F_\lambda(\vec r).
\end{align}  
The resulting source operator reads as 
\begin{align}
   \hat{S}(t)&=\frac{\omega_\text s}{\hbar}\int dt'\theta(t-t')e^{-i\omega_\text s(t-t')}\vec A_\text s(\vec r_\text s)\cdot\hat{\vec C}(t').
\end{align}
All time dependences are encoded in this operator, while the spatial structure is encoded in $\vec\chi(\vec r)$. Eventually, Eq.~(\ref{eq.sourcefields}) follows from projection onto the propagation direction.

\end{document}